\documentclass[conference]{IEEEtran}
\IEEEoverridecommandlockouts
\usepackage{cite}
\usepackage{amsmath,amssymb,amsfonts}
\usepackage{algorithmic}
\usepackage{graphicx}
\usepackage{textcomp}
\usepackage{xcolor}

\usepackage{todonotes}
\usepackage{color}

\usepackage{comment}

\def\BibTeX{{\rm B\kern-.05em{\sc i\kern-.025em b}\kern-.08em
    T\kern-.1667em\lower.7ex\hbox{E}\kern-.125emX}}
\begin{document}

\title{IEEE 7010: A New Standard for Assessing the Well-being Implications of Artificial Intelligence}
%{\footnotesize \textsuperscript{*}Note: Sub-titles are not captured in Xplore and should not be used}
%\thanks{This work is connected to an IEEE working group on Well-being Metrics for Autonomous and Intelligent Systems (P7010 Standard)}
%}

\author{\IEEEauthorblockN{1\textsuperscript{st} Daniel Schiff}
\IEEEauthorblockA{\textit{}
\textit{Georgia Institute of Technology, }\\
Atlanta, USA  \\
ORCID: 0000-0002-4376-7303}
\and

\IEEEauthorblockN{2\textsuperscript{nd} Aladdin Ayesh}
\IEEEauthorblockA{\textit{} %Faculty of Computing, Engineering and Media
\textit{De Montfort University, }\\
Leicester, UK \\
ORCID: 0000-0002-5883-6113}
\and 

\IEEEauthorblockN{3\textsuperscript{rd} Laura Musikanski}
\IEEEauthorblockA{\textit{} 
\textit{Happiness Alliance}\\
Seattle, USA \\
ORCID: 0000-0002-7477-6683}
%\and
%\IEEEauthorblockN{4\textsuperscript{th} Zvikomborero Murahwi}
%\IEEEauthorblockA{\textit{dept. name of organization (of Aff.)} \\
%\textit{name of organization (of Aff.)}\\
%City, Country \\
%email address or ORCID}
\and

\IEEEauthorblockN{4\textsuperscript{th} John C. Havens}
\IEEEauthorblockA{\textit{The IEEE Global Initiative on }\\ 
\textit{Ethics of Autonomous and Intelligent Systems }\\
Piscataway, USA \\
ORCID: 0000-0002-7226-422X}
%\and
%\IEEEauthorblockN{6\textsuperscript{th} Given Name Surname}
%\IEEEauthorblockA{\textit{dept. name of organization (of Aff.)} \\
%\textit{name of organization (of Aff.)}\\
%City, Country \\
%email address or ORCID}
}

\maketitle

\begin{abstract}
Artificial intelligence (AI) enabled products and services are becoming a staple of everyday life. While governments and businesses are eager to enjoy the benefits of AI innovations, the mixed impact of these autonomous and intelligent systems on human well-being has become a pressing issue. This article introduces one of the first international standards focused on the social and ethical implications of AI: The Institute of Electrical and Electronics Engineering’s (IEEE) Standard (Std) 7010-2020 Recommended Practice for Assessing the Impact of Autonomous and Intelligent Systems on Human Well-being. Incorporating well-being factors throughout the lifecycle of AI is both challenging and urgent and IEEE 7010 provides key guidance for those who design, deploy, and procure these technologies. We begin by articulating the benefits of an approach for AI centered around well-being and the measurement of well-being data. Next, we provide an overview of IEEE 7010, including its key principles and how the standard relates to approaches and perspectives in place in the AI community. Finally, we indicate where future efforts are needed.\\
\end{abstract}

\begin{IEEEkeywords}
Well-being, artificial intelligence, autonomous systems, measurement, design
\end{IEEEkeywords}

\section{Introduction}
Current growth in AI-enabled applications and products is facilitating the integration of technology into the fabric of modern life \cite{makridakis_forthcoming_2017}. This innovation is manifesting in a wide range of daily uses---from smart devices and industrial robotics ---and promises to continue as new use cases are developed in healthcare \cite{yu_artificial_2018}, finance \cite{bahrammirzaee_comparative_2010}, transportation \cite{fagnant_preparing_2015}, and throughout commercial, domestic, and the public spaces \cite{reisman_algorithmic_2018}. Autonomous and intelligent systems\footnote{IEEE 7010 uses the term “autonomous and intelligent systems” (A/IS or AIS) to describe what is termed in this article “artificial intelligence” (AI).} are central to a sea change, as they are increasingly ubiquitous, impacting the public in visible and invisible ways \cite{schuster_ambient_2007}. We propose that this sea change has a profound potential to impact human physical and mental well-being in both positive and negative ways.

Well-being is important to overall quality of life \cite{diener_subjective_1984,huppert_flourishing_2013,odonnell_wellbeing_2014}, and while some research has been done on well-being in the digital era \cite{diener_subjective_1999, huang_internet_2010,liu_digital_2019,twenge_decreases_2018,calvo_positive_2014}, measuring the impact of AI on users as well as those indirectly affected is still in its early stages \cite{oecd_artificial_2019, musikanski_artificial_2020}. Issues such as algorithmic bias and lack of transparency in facial recognition, natural language processing, and criminal justice \cite{chouldechova_case_2018,buolamwini_gender_2018} have captured public interest, along with the role of AI in targeted advertising and political misinformation \cite{little_fake_2018,chesney_deepfakes_2019}. Privacy \cite{tene_big_2012}, inequality \cite{piketty_capital_2014}, social cohesion \cite{west_future_2018}, and labor displacement \cite{autor_why_2015} are among the many social and ethical risks to human well-being associated with AI. In response, governments, corporations, and NGOs have proposed new ethical codes, principles, frameworks, industry standards, and policy strategies \cite{jobin_global_2019,schiff_whats_2020,musikanski_ieee_2018}. Assessments of AI’s impact on human and societal well-being are relatively new, and there is not much known about how to incorporate a well-being orientation and measurement into organizational or public settings. Thus, we suggest that IEEE 7010 offers initial guidance and sets the stage for increasingly robust and rigorous development, deployment, and evaluation of AI in the future.

In this article, we review the first industry standard to our knowledge that addresses the well-being implications of AI.\footnote{The authors of this paper were members of the working group that formed IEEE 7010 standard as well as a chapter of IEEE's Ethically Aligned Design, First Edition focused on well-being, published in March 2019 (The IEEE Global Initiative on Ethics of Autonomous and Intelligent Systems 2019). This article does not represent an official position of IEEE, but rather the opinions of its authors.} The Institute of Electrical and Electronics Engineering (IEEE) Standard 7010-2020 Recommended Practice for Assessing the Impact of Autonomous and Intelligent Systems on Human Well-being (henceforth IEEE 7010) is part of the IEEE 70xx series, a series of standards designed to address ethical dimensions of AI \cite{chatila_ieee_2019, koene_ieee_2019}. Projects in the IEEE 70xx series include IEEE P70xx series standards on ethics in design (IEEE P7000), transparency (IEEE P7001), data privacy (IEEE P7002), algorithmic bias (IEEE P7003), child and student data (IEEE P7004), employer data governance (IEEE P7005), data agents (IEEE P7006), ontologies for ethics (IEEE P7007), nudging (IEEE P7008), fail-safe design (IEEE P7009), trustworthiness of news (IEEE P7011), machine readable privacy (IEEE P7012), facial recognition (IEEE P7013), and empathy (IEEE P7014). 
IEEE 7010 is oriented around a holistic well-being perspective and offers practical guidance for AI creators seeking to understand and measure direct, indirect, intended, and unintended impacts to human and societal well-being.

In Section II we argue for the urgency of well-being in the context of increasing uses of AI and describe the benefits of a well-being perspective. We suggest that such an approach to AI development and measurement can help organizations build awareness, provide actionable evidence and insights, develop infrastructure, manage risks, and improve well-being of individuals and groups in society. 

Section III provides an overview of the core principles and processes that constitute IEEE 7010, including its well-being impact assessment (WIA). In this section, we describe how IEEE 7010 complements and differs from other social and ethical orientations to assess the impacts of AI, such as human rights; fairness, accountability, and transparency (FAT); law; and sustainability. We also explain how IEEE 7010 could help governments in the formation of regulations for AI. 

Section IV offers considerations for how IEEE 7010  can be applied within an organizational context. We propose that implementing a well-being approach in alignment with traditional software and engineering lifecycle processes could take a few different forms. We explore the application of well-being as a principle, incorporated into system design as metrics, and via lifecycle development stages. We present hypothetical case studies for autonomous vehicles and healthcare robots to help demonstrate how IEEE  7010 might be applied fruitfully to prominent AI use cases.  \\

\section{Motivation}

\subsection{The urgency of well-being in the autonomous/intelligent systems era}
Well-being is a complex, multidimensional, and occasionally contested subject studied  for almost half a century by scholars in psychology, management, economics, and policy \cite{diener_subjective_1999, easterlin_income_2001, stiglitz_measuring_2019}.  Some scholars of well-being distinguish between hedonic, eudaimonic, and social dimensions of well-being. Hedonic well-being emphasizes immediate subjective experiences, such as positive and negative affect and life satisfaction (National Research Council 2013), while the eudaimonic dimension of well-being takes into account a broader definition of flourishing \cite{fisher_conceptualizing_2014}. Well-being also has both objective and subjective components, where the former refers to traditional economic metrics like health and income, and the latter to internal subjective experience \cite{alatartseva_well-being_2015, dale_subjective_1980}. The Organization for Economic Co-operation (OECD) (2011; 2013) conceives of well-being as incorporating numerous domains of social and economic life such as education, economy, environment, health, government, community, culture, work, psychological well-being, and human settlements and encompassing both subjective and objective components.  Well-being incorporates an orientation towards human rights, one which enjoys wide support and is deeply grounded in legal systems \cite{latonero_governing_2018}, as well environmental sustainability, now maturely incorporated into industry standards and processes \cite{bansal_strategic_2003}. Over time, the study of well-being has matured into an increasingly comprehensive, theoretically and empirically rigorous, and applied field, that is gaining traction among researchers  as a perspective and in public policy \cite{odonnell_wellbeing_2014, stiglitz_measuring_2019}.

A focus on well-being can offer an alternative approach to a traditional focus on economic growth and financially-driven efficiency. While the obligations of corporations to the public have long been a topic of discussion \cite{campbell_why_2007,ruder_public_1965}, concerns about corporate social responsibility have grown in recent years due to evidence of persistent economic inequality and ecological degradation \cite{piketty_capital_2014,wilkinson_inner_2020,wilkinson_spirit_2009,stiglitz_report_2009,hawken_natural_2013}. Stout \cite{stout_shareholder_2012} as well as Tepper and Hern \cite{tepper_myth_2018} emphasize recent trends in corporate governance, arguing that corporations which prioritize short-term financial gain too often fail to protect the public. These concerns have given rise to calls for increased corporate social responsibility and governmental action that protect human and societal well-being. Movements focused on corporate social responsibility that incorporate environmental sustainability and human well-being \cite{miles_environmental_2000, armstrong_effects_2013, brown_rise_2009} are gaining purchase as a variety of new conceptual and organizational strategies have surfaced. These include the Beyond GDP \cite{kubiszewski_beyond_2013}, triple bottom line \cite{elkington_partnerships_1998}, and social enterprise models \cite{dart_legitimacy_2004}, now enhanced via formal certification of socially responsible organizations as B Corps \cite{stubbs_sustainable_2017} and through legal incorporation as public benefit corporations \cite{hiller_benefit_2013}. 

National and intergovernmental bodies have also increasingly broadened their focus beyond simple economic growth, such as through the United Nations’ Sustainable Development Goals (SDGs), a landmark set of international goals and strategies to achieve poverty reduction, improved health, sustainability, and other aspects of sustainable development that include dimensions of well-being \cite{united_nations_transforming_2015, iriarte_bridging_2019, de_neve_world_2020}. Meanwhile, the measurement of national well-being, such as the gross national happiness (GNH) index \cite{ura_bhutan_2012}, and the United Kingdom’s Office of National Statistics Well-being measures \footnote{See: www.ons.gov.uk/peoplepopulationandcommunity/wellbeing}, has now moved from national to international prominence \cite{stiglitz_measuring_2019, musikanski_happiness_2014}. Well-being has become a mainstream topic in international governance bodies like the OECD and United Nations \cite{diener_well-being_2009} and the World Happiness Report now evaluates well-being in 156 countries \cite{helliwell_world_2020}. Engineering and computing have also increased calls for responsible research and innovation (RRI) \cite{schomberg_vision_2013} in design processes, incorporating new tools, methodologies, and perspectives surrounding ethical design \cite{dignum_ethics_2018}.

This evolving understanding of the nature of human and societal well-being is occurring at the same time that we are experiencing unprecedented technological transformation. The increased influence of technology companies, recent concerns over privacy and security \cite{chesney_deep_2018}, 
propaganda \cite{chesney_deepfakes_2019}, algorithmic bias %\cite{bolukbasi_man_2016}
, and labor displacement \cite{frey_future_2013} 
have put the power of autonomous and intelligent systems into the limelight, adding to prior concerns about military uses of AI \cite{marchant_international_2015} 
and long-term safety \cite{muller_future_2016}. 
Recent applications of AI to manage exposure to Covid-19 have also raised privacy concerns. \cite{nanni_give_2020}.

Scholars, policymakers, and the public now widely recognize that AI, as a general purpose technology, has an enormous scope and scale of potential impact across every sector of society. Schwab \cite{schwab_fourth_2016} has argued that AI is central to an impending fourth industrial revolution, likely to transform industry and public life over the next decades. The McKinsey Global Institute \cite{bughin_modeling_2018} (2018)  estimated AI’s financial impact at trillions of dollars in the near future. We suggest that there is an urgent need to consider the sweeping implications of AI. We propose that it is critical that the design, deployment, evaluation, and regulation of AI prioritize the measurement and safeguarding of well-being in society. We further suggest that a holistic approach to well-being is useful for understanding and managing the impacts of AI on human and societal well-being. 

\subsection{A holistic approach to AI’s impacts on well-being}
A holistic and impact-based approach to well-being, widely defined, differs from approaches based on technological solutions alone. Tools and methodologies centered on technical fixes can fail to take account of the full social context and set of ethical issues surrounding AI \cite{selbst_fairness_2018}. They may emphasize design processes presumed to promote ethics, while failing to actually assess impacts. Moreover, technical fixes can also assign responsibility for anticipating and reacting to well-being considerations too narrowly, focusing mostly on the role of engineers and computer scientists \cite{greene_better_2019} while minimizing corporate responsibility. 

Technical solutions and strategies do have an important role in ensuring responsible design and use of AI, and the understanding of how these methodologies work is growing \cite{morley_what_2019}. However, much more needs to be understood about the fields of AI and well-being in both industry and other sectors (governmental, educational, etc.). Moreover, we doubt that technical fixes alone will ever adequately meet challenges that are inherently both technical and social in nature and so we propose that incorporating both social and technical elements of AI into standards is a better approach to  understanding, measuring, and managing the direct, indirect, intended, and unintended impacts of AI on human and societal well-being.

In short, well-being is a robust and promising orientation for considering the human and social and ethical implications of AI. We propose that organizations have much to gain by adopting this perspective. What is needed is a framework to help organizations go about the process of adoption and adaptation of a well-being framework for understanding and managing AI impacts on human and societal well-being. 

\subsection{Benefits of a well-being orientation}

The IEEE 7010 standard is designed for AI creators, including organizations that wish to design, deploy, procure, or evaluate these systems for their impact on the well-being of humans. An orientation to AI that revolves around a well-being perspective provides five benefits:
\begin{itemize}
    \item \textbf{Building awareness}: building awareness is a critical first step to organizational change. Diverse organizations and roles can come to understand the importance and relevance of well-being to their work.
    \item \textbf{Providing evidence}: measuring well-being provides concrete data regarding impacts in place of ignorance or assumptions that certain impacts are inevitable or unlikely. Measurement allows organizations to evaluate successes and failures, which can be used in goal setting, prioritization, and management.
    \item \textbf{Developing infrastructure}: incorporating well-being helps organizations build the necessary infrastructure for thinking about, analyzing, and responding to well-being data. Developing infrastructure includes processes for integrating well-being considerations into data collection, decision-making, identifying roles and responsibilities, and business use case development, all areas that require organizational infrastructure.
    \item \textbf{Managing risks}: well-being measurement and reporting helps organizations and society both individually and collectively assess the risks of various AI products and services. Managing risks safeguards employee, customer, user, and public well-being, protects against reputational and brand threats, and aids in anticipating and responding to regulatory and legal requirements.
    \item \textbf{Improving well-being}: awareness, evidence, infrastructure, and risk management jointly enable organizations to convert learning about well-being into positive improvements to AI design. Improvements to well-being enhance brand quality and trust, safeguard against harms to well-being, and enable the realization of new opportunities to enhance well-being.
\end{itemize}

Below, we introduce IEEE 7010.\\

\section{IEEE 7010: A New Standard for Assessing the Well-being Implications of AI}
\subsection{Introduction to IEEE 7010}
Importantly, IEEE standards are developed by working groups of volunteers, often individuals from many sectors of society from around the world. Ther working group members of IEEE 7010 had expertise in AI, well-being, organizational science, policy, psychology, and many other areas. Standards development often requires several years, including multiple rounds of external review and voting before a standard is approved. IEEE 7010 is one of the standards in the IEEE P70xx series of standards that emerged from a focus on ethical dimensions of AI.

IEEE issues four classifications of standards: standards, recommended practices, guidance, and best practices. IEEE 7010 is a recommended practice, which identifies recommended procedures that adherents to the standards should apply. A recommended practice is stronger than a standard classified as a guide or trial-use document, which states how adherents can apply a certain practice, but less stringent than a standard of the classification of standard, which states how adherents must comply. The status of IEEE 7010 as a recommended practice reflects the state of the field and the need to develop an understanding of the impact of AI on human and societal well-being, such as through an Well-being Impact Assessment (WIA), especially within organizational settings. 

\begin{figure}[h!]
	\centering
	\includegraphics[width=0.5\textwidth]{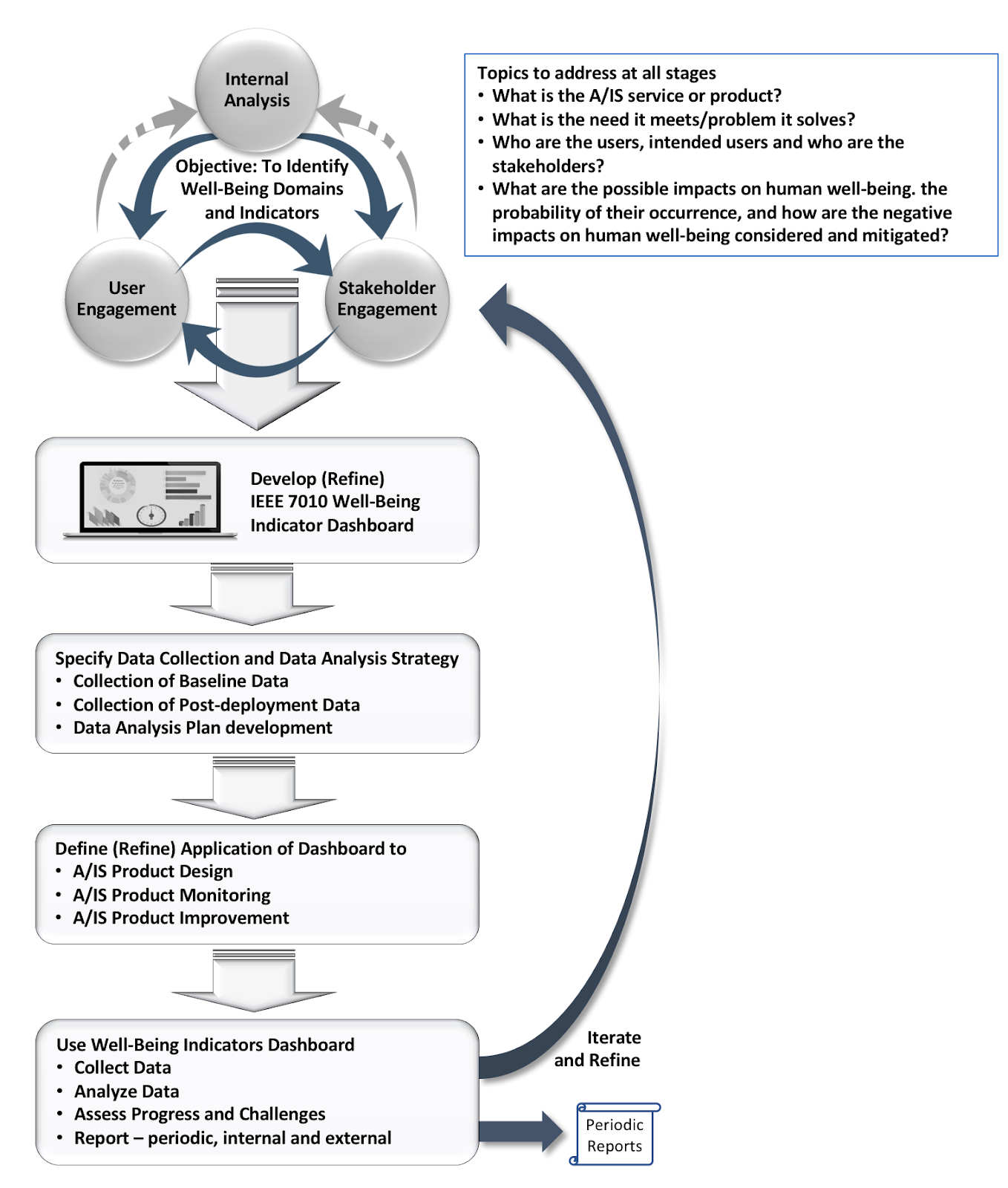} 
	\caption{Flowchart of IEEE 7010 WIA (IEEE P7010 - Adapted and reprinted with permission from IEEE. Copyrights IEEE 2020. All rights reserved.)}
	\label{fig1:7010flow}
\end{figure}

As with all IEEE standards, IEEE 7010 describes its scope, purpose, and key definitions. IEEE 7010 also contains annexes to support the core of the standard, including examples of well-being indicators, examples of particular AI use cases, examples of hypothetical organizations applying parts of the standard, guidance on managerial adoption, a discussion of the value of IEEE 7010 to a wide range of stakeholders, and other useful resources and references. The annexes and introduction provide additional guidance around the core components of IEEE 7010, the WIA. The WIA focuses on five activities, each broken down into several tasks, as depicted in \ref{fig1:7010flow}. These activities and tasks are grounded in stakeholder engagement and an iterative process.

\subsubsection*{Activity 1. Internal analysis and user and stakeholder engagement}
The first activity of the WIA focuses on internal evaluation of an AI product or service and information gathering from users and broader stakeholders regarding its possible impacts. In terms of the internal analysis task, organizations should answer five questions involving 1) the nature of the AI system, 2) the needs it meets or problems it solves, 3) who the users (intended and unintended) are, 4) who the broader stakeholders might be, and 5) the likelihood of possible positive and negative impacts, and how can they be considered and mitigated. The internal analysis task allows organizations to have a broad and thoughtful conversation about the purpose and impacts of its AI. This goes beyond considering user experience, as it requires considering the full range of possible impacts for users and stakeholders. (Stakeholders are defined as any individuals or groups who are or might be affected by the AI.)

The initial assessment of AI’s possible impact is not done only internally. Organizations engage directly with users and stakeholders, such as through interviews and focus groups. These tasks include asking users and stakeholders about how they use or intend to use the AI, and what possible impacts, benefits, and harms they experience or anticipate. The user and stakeholder engagement tasks of IEEE 7010 asks that organizations go well beyond typical user experience and product testing processes, and evaluate the full implications of the AI across its use and lifecycle. For example, organizations that develop AI products for K-12 education should expect to engage with students, teachers, parents, administrators, and potentially other stakeholders. An organization that sells AI products to government entities for use in the criminal justice system should be prepared to engage with government employees and individuals in the criminal justice system. 

User and stakeholder engagement and internal analysis are part of a continual reflective process, where multiple iterations may be performed as the organization gets a better understanding of its AI and its well-being impacts. Importantly, this first activity and three associated tasks are centered at acquiring evidence surrounding the well-being impacts of a specific AI product or service. As such, organizations should be deliberate about a wide range of possible impacts. To do so, they look to the well-being indicators described in IEEE 7010. The role and selection of well-being indicators is described below.\\

\subsubsection*{Activity 2. Development and refinement of well-being indicators dashboard} IEEE 7010 identifies indicators of human and societal well-being across multiple dimensions of well-being, divided into twelve domains. These domains are affect, community, culture, education, economy, environment, health, human settlements, government, psychological/mental well-being, and work. IEEE 7010 provides sample indicators associated with each domain as well as additional resources through which to identify additional indicators. The well-being indicators within IEEE 7010 are sourced from mainstream evaluative tools for well-being, and are validated by rigorous research. This is a hallmark feature of IEEE 7010.

\begin{figure}[h!]
	\centering
	\includegraphics[width=0.5\textwidth]{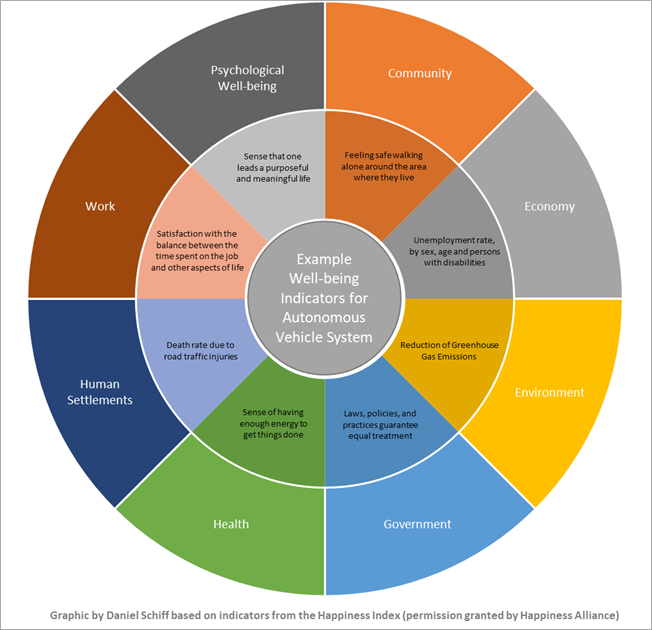} 
	\caption{Sample well-being indicators dashboard for autonomous vehicles}
	\label{fig2:AVdashboard}
\end{figure}

The internal analysis and user and stakeholder engagement tasks center around providing evidence for identifying possible well-being indicators that measure the impact of the AI. Users and stakeholders are asked about impacts to their environment, health, work, community, social support and other dimensions of well-being. The indicators and domains guide the creation of a well-being indicators dashboard. The dashboard can be represented visually as depicted in \ref{fig2:AVdashboard} and integrated into engineering processes. 

During the dashboard creation process, organizations should articulate: 1) the source for each indicator, 2) why the indicator was selected, 3) the associated domain from the list of twelve, and 4) if appropriate, any adaptations made to indicators. IEEE 7010 recognizes that existing indicators of well-being may not adequately reflect the impact of a given organization’s AI product or service on human or societal well-being. For example, some indicators focus on measurement of an aspect of well-being at the national level, whereas an AI may impact a smaller number of individuals. IEEE 7010 allows for customization and flexibility when appropriate, while aiming to stay as close as possible to scientifically valid well-being indicators.\\

\subsubsection*{Activity 3. Data planning and collection}
As IEEE 7010 is at heart a WIA, data collection is important. The third activity follows the development or refinement of the well-being indicators dashboard. Based on the identified indicators and domains, an organization articulates a data collection plan. This includes collection of both baseline data and data over time, allowing changes in well-being indicators to be assessed over time. The data collection plan task involves organizations describing what data will be collected and how. This includes identifying sources (e.g., which users or stakeholders), methods (e.g., surveys, product metrics), and the frequency and timelines associated with data collection.

After establishing a data collection plan, the next task is to collect the data. This includes collecting baseline data for users and stakeholders as well as populations that are statistically similar to users and stakeholders. Collecting data on the latter group helps to set a baseline and is useful for comparative analysis. Next, data are collected over time, after users and stakeholders engage with a given AI product or service. Collecting data from multiple populations and over time allows for assessment of actual impacts. IEEE 7010 recognizes that collecting data and associating changes over time with a particular AI product or service is conceptually and empirically challenging. However, the challenge of identifying specific causal impacts should not lead to the abandonment well-being impact assessment, but instead to refine and improve it.\\

\subsubsection*{Activity 4. Data analysis and improvement to AI}
The fourth activity is analysis and use of collected data. Organizations can identify trends over time and changes with respect to baseline data or compared to other populations. Data analysis can also illuminate unexpected uses, behaviors, and impacts, and to document how a particular AI product or service was used in real-world settings. Analysis helps determine if an AI does have negative impacts, or if efforts to mitigate negative impacts or increase positive impacts are successful. Importantly, analysis then feeds into improvements to AI design, development, assessment, monitoring, and management. IEEE 7010 does not specify all of the actions that an organization should take, as how an organization learns from the WIA and data depends on the specific product or service, its impacts, and organizational circumstances. However, there are some actions IEEE 7010 recommends towards refining well-being impact assessment, described in activity five.\\

\subsubsection*{Activity 5. Iteration}
IEEE 7010 presents activities that should be performed in an iterative fashion. There are deliberate feedback cycles built in. Assessment of the well-being impacts of an AI product or service should not be a one-time effort and should be ongoing. Organizations can use lessons learned from the user and stakeholder engagement  process and data analysis to improve the implementation of the well-being assessment process. They should also refine the well-being indicators dashboard as their understanding grows. Improving the assessment process also means strengthening data collection and analysis plan. Over time, organizations should develop capacity towards streamlining WIA, learning how to conceive of impacts more robustly, capture and analyze data more accurately, and efficiently implement changes into engineering, software, and business processes. Finally, organizations can report both internally and externally to users and stakeholders, in order to help assess and communicate progress and opportunities for improvement.\\

\subsubsection*{Note on Data Collection and Privacy}
Data collection is at the heart of WIA. Collected data may be in the form of user surveys, publicly available government sources, or usage data, for example, mouse clicks \cite{renaud_making_2004}, health/physiological data \cite{li_digital_2017}, and facial expressions \cite{saneiro_towards_2014}. However, these and other forms of data are all potentially highly sensitive based on how data are governed, secured, and used. In light of the rising prospect of cognitive profiling through affect state estimation or personalized analytics \cite{ayesh_towards_2016}, privacy becomes of urgent importance.

In response, several policy-making institutions have produced guidelines, code of conducts, or model laws to address the challenge to individual privacy and data usage rights arising from the rapid development of data-intensive AI and other applications \cite{bennett_governance_2017}. Among the most extensive laws is the European Union's (EU) General Data Protection Regulation (GDPR)\footnote{EU: See: https://eugdpr.org}, which applies to all EU member countries and is now referenced by other nations \cite{chander_catalyzing_2019}. The EU's GDPR is often considered a baseline for data protection laws \cite{voigt_eu_2017}. There is a potential tension between data collection for well-being improvement and user privacy, though there are privacy-preserving approaches \cite{nanni_give_2020}. IEEE 7010 is not a standard for data protection or other considerations such as algorithmic bias or data governance but includes mention of protective measures such as the EU's GDPR as well as field- or sector-specific guidelines.  \\
 
\section{Relationship of IEEE 7010 to Other Processes and Approaches}
IEEE 7010 complements, extends, and differs from other orientations, tools, and methodologies. Below we describe its relationship to some prominent approaches relevant to assessing the social and ethical impacts of AI.\\

\subsubsection{The IEEE 70xx standards series} The IEEE Global Initiative on Ethics of Autonomous and Intelligent Systems is among the earliest and most robust efforts, resulting in three drafts of IEEE’s Ethically Aligned Design issued in 2016, 2017, and 2019) \cite{the_ieee_global_initiative_on_ethics_of_autonomous_and_intelligent_systems_ethically_2019} and a proposed a series of 14 standards. The standard series addresses AI ethics issues including privacy, bias, transparency, trustworthiness of news, empathy, fail-safe design, ontologies, and ethical design \cite{chatila_ieee_2019}. Because at the time of publishing of IEEE 7010, other standards in the IEEE 70xx series had not been published, it does not recommend adherence to other standards.). However, it can be anticipated that other IEEE 70xx standards will be relevant to the scope and purpose of Std 7010, and future revisions of IEEE 7010 may incorporate these standards.\\

\subsubsection{Sustainability} While IEEE 7010 includes “Assessing the Impact of Autonomous and Intelligent Systems on Human Well-being” in the title, the standard explicitly includes sustainability in the definition of well-being, especially with respect to the environment. IEEE 7010’s definition of well-being is “the continuous and sustainable physical, mental, and social flourishing of individuals, communities and populations where their economic needs are cared for within a thriving ecological environment” (IEEE 7010 2020). The environment domain in IEEE 7010 considers many possible impacts of AI, such as on pollution, waste, and ecological diversity. Stakeholders concerned with environmental and ecological sustainability and well-being should view IEEE 7010 as a supportive tool.\\

\subsubsection{Human rights} Human rights is an important aspect of human and social well-being. Human rights have a long tradition in legal philosophy and international governance. It has been applied to AI recently by scholars \cite{cath_leap_2020, latonero_governing_2018} and IEEE’s Ethically Aligned Design includes human rights as indispensable (The IEEE Global Initiative on Ethics of Autonomous and Intelligent Systems 2019). IEEE 7010 is consistent with and embraces a human rights perspective. It incorporates assessment of human rights indicators within the domain of government.\\

\subsubsection{Fairness, accountability, and transparency} Fairness, accountability, transparency, and ethics are referred to as FAT, FATE, FEAT, or similar acronyms. These approaches typically emphasize ethical issues in algorithm design that are seen as technically tractable, especially algorithmic bias and transparency \cite{mittelstadt_explaining_2019, rakova_assessing_2020}. This approach is popular amongst computer science and AI ethics researchers, such as the Academy of Computing Machinery’s FAccT conference and community. Recently, these communities are opening up to broader perspectives, such as social science, law, and policy \cite{selbst_fairness_2018}. 
Scholars and practitioners interested in this perspective may find that IEEE 7010 can serve as an important extension to their toolset, one that allows them to conceive of AI’s impacts on human and societal well-being more broadly. For example, practitioners of FAT can apply IEEE 7010’s WIA and then apply FAT tools and methodologies based on impacts through the WIA and data. \\

\subsubsection{Algorithmic impact assessments} Algorithmic impact assessments are sometimes focused on the downstream effects of AI as well as ethical issues during the design stage, as they often heavily feature aspects of FAT. These assessments are promoted by scholars and nongovernmental organizations (NGOs) \cite{Calvo_Peters_Cave_2020, reisman_algorithmic_2018}, and are being considered as part of regulatory frameworks. The European Commission proposed elements of algorithmic assessment within its most recent guidance for the EU \cite{european_commission_artificial_2020, kaminski_algorithmic_2019}, and the United States has proposed the Algorithmic Accountability Act \cite{quezada_tavarez_algorithmic_2019}. IEEE 7010 is highly concordant with this perspective, as it is centered on impact assessment. IEEE 7010’s WIA could serve as a tool for policymakers to look to in the near future. \\

\subsubsection{Law and regulation} IEEE 7010 is a voluntary industry standard, to be considered as part of private sector self-governance or collective governance. Other standards, such as those created by national standards bodies, may be legally mandatory. However, voluntary standards can play a role in legal frameworks and have political legitimacy \cite{bernstein_can_2007, kolk_effectiveness_2002}. For example, adherence to an industry standard can demonstrate good faith efforts beyond minimal levels of compliance and can reduce a company’s liability. Moreover, governments and courts of law sometimes give consideration to companies adhering to industry standards when contracting or grant making. For example, the 2016 U.S. National Research and Development (R\&D) Plan states that “Industry and academia are the primary sources for emerging AI technologies” and considers “Updating acquisition processes across agencies to include specific requirements for AI standards in requests for proposals” \cite{national_science_and_technology_council_national_2016}. IEEE 7010 may provide a means for companies to demonstrate readiness to adhere to future laws and regulations.\\

\subsubsection{Software and engineering processes} The incorporation of well-being measurement into traditional and newer software engineering processes, such as agile software development and engineering lifecycle standards \cite{laporte_software_2008} is critical to organizations that develop AI, and remains a challenge \cite{rakova_assessing_2020}. IEEE 7010 includes a description of its application to the Plan-Do-Check-Act Cycle, Lifecycle Analysis, and other processes in its Annexes.  IEEE 7010 features checklists throughout each of its activities and tasks within the well-being assessment, consistent with checklist-based processes are common in engineering design \cite{madaio_co-designing_2020}. %Currently, we are investigating three approaches to achieve integration into software and engineering processes, which  are based on translating the recommended practices into (1) a set of principles, (2) a set of measurable metrics, or (3) a set of processes to be incorporated in the product lifecycle stages. The result of this future research is to be reported in a follow-up publication.\\

\section{Recommendations}
IEEE 7010 is the first international standard for assessing and managing the impact of well-being on humans and society from AI. As such, there are not yet examples or case studies of its use. We recommend the following eight areas for development: \\

\begin{enumerate}
    \item \textbf{Use of IEEE 7010.} AI creators, managers and investors use IEEE 7010. We recommend the use of IEEE 7010 to assess, manage, mitigate, and improve the well-being impacts on human and societal well-being, extending from individual users to the public. This recommendation covers AI projects in their inception to those fully implemented in the marketplace. This recommendation is equally important for independent or small projects as for large-scale projects undertaken by international or transnational organizations. 
    
    \item \textbf{Development of organizational structures and processes.} AI creators, managers, and investors develop processes, tools, and resources for implementing and maintaining IEEE 7010 into their organizational processes and structures. This includes developing staffing and workflow structures, engineering and computing practices, hiring and training, internal resources and toolkits, external communication, and other necessary structures.

    \item \textbf{Formation of an Industry Connections group.} IEEE and industry stakeholders invested in well-being and AI form an Industry Connections (IC) group\footnote{https://standards.ieee.org/industry-connections/index.html}. An IC group creates an ecosystem for parties to perform research, create tools and resources, publish information, and promote its work via conferences and other events. We recommend that such an IC group be created to help translate the IEEE 7010 standard into practice.
    
    \item \textbf{Shared language of well-being.} Researchers and practitioners in the AI or well-being field define well-being broadly to encompass the multi-dimensional aspects of well-being that includes subjective and objective measurements included in IEEE 7010 across the domains of: “(1) Affect, (2) Community, (3) Culture, (4) Education, (5) Economy, (6) Environment, (7) Human Settlements, (8) Health, (9) Government, (10) Psychological Well-Being/Mental well-being, (11) Satisfaction with life and (12) Work” \cite{musikanski_ieee_2018}. To encourage consensus-building, we further recommend that researchers and practitioners that focus on a single dimension or subset of dimensions of well-being identify their scope of study within this multidimensional definition of well-being.
    
    \item \textbf{Expanded understanding of well-being.} AI creators, managers and investors develop an understanding of the definition of human and societal well-being conceptually and in terms of measurement and data. They should develop a deeper understanding of the impacts on human and societal well-being from AI in general and related to their specific AI work. We further recommend that leading individuals and organizations in the AI field bring conversation about the impacts on human and societal well-being from AI to the forefront through publications, conferences, competitions, and other means.
    
    \item \textbf{Increased research on AI and well-being.} Researchers and practitioners develop the field of knowledge about the impacts of AI on human and societal well-being gathering and analyzing well-being data from users and stakeholders of specific AIs, as well as the general population. We further recommend that AI creators, managers, and investors cooperate with researchers in the gathering and analysis of well-being data and that researchers and practitioners cooperate with each other through sharing of data. We also recommend that researchers, practitioners, and AI creators jointly contribute to the formation of best practices, such as to evolve IEEE 7010 and similar efforts.
    
    \item \textbf{Educational efforts focused on well-being.} Universities develop courses, centers, research programs, and other means of teaching and developing skills for future AI creators, managers, and investors. We recommend that professors and instructors currently teaching AI integrate IEEE 7010 into their curriculum and that university and other educational institutions develop programs and courses which integrate IEEE 7010, with the aim of developing future AI creators’, managers’, and investors’ capacity to integrate well-being metrics and goals into their future work.
    
    \item \textbf{Regulation of AI’s impacts on well-being.} Governments formulate regulations and standards to contribute to securing and protecting the well-being of humans and society in light of the impacts of AI. We recommend that governments collaborate in the formulation of such standards or regulations as well as develop these independently.  We also recommend that international non-governmental organizations facilitate this effort through promulgation of model codes and regulations, as well as issuance of proclamations and resolutions reinforcing their importance, and the coordination of efforts at high-level meetings, summits, and conferences aimed. We recommend that governments adopt a process similar to or based on IEEE 7010 for any development of or investment in AI. Finally, we recommend that governments consider adoption of IEEE 7010 as part of grant making procedures and to demonstrate regulatory compliance.\\
\end{enumerate}

\section{Conclusion}
This article introduced IEEE 7010, a new international standard aimed at addressing the well-being implications of AI. After describing the concept of well-being, we identify the potential benefits of a well-being orientation in light of the social and ethical issues of the modern world. Next, we provide an overview of IEEE 7010, outlining the core activities and tasks that define its assessment of well-being. We then situated IEEE 7010 in relation to other processes and approaches currently used to think about social, ethical, and well-being implications related to AI. Finally, we offered recommendations for a variety of stakeholders. \\

\begin{comment}
\section*{Acknowledgment}

The preferred spelling of the word ``acknowledgment'' in America is without 
an ``e'' after the ``g''. Avoid the stilted expression ``one of us (R. B. 
G.) thanks $\ldots$''. Instead, try ``R. B. G. thanks$\ldots$''. Put sponsor 
acknowledgments in the unnumbered footnote on the first page.
\end{comment}

\bibliographystyle{IEEEtran}

\bibliography{ieeesmc}

\end{document}